\documentclass[preprint,aps,nofootinbib]{revtex4}
\usepackage{float}
\usepackage{graphicx}
\input{epsf.sty}
\usepackage{epsfig}

\def\beq{\begin{equation}}
\def\eeq{\end{equation}}
\def\lsim{\mathrel{\raise.3ex\hbox{$<$\kern-.75em\lower1ex\hbox{$\sim$}}}}
\def\gsim{\mathrel{\raise.3ex\hbox{$>$\kern-.75em\lower1ex\hbox{$\sim$}}}}

\begin{document}

\vskip 0.4cm
\title{ Charge effect and  finite 't Hooft coupling correction
 on drag force and Jet Quenching Parameter }
\author{ Kazem Bitaghsir Fadafan\\Physics Department, Shahrood University of
Technology,\\ P.O.Box 3619995161 Shahrood, IRAN\\
bitaghsir@shahroodut.ac.ir}
%\date{\today}
\vspace*{2.0cm}
\begin{abstract}
The effects of charge and finite 't Hooft coupling correction on
drag force and jet quenching parameter are investigated. To study
charge effect and finite 't Hooft coupling correction , we consider
Maxwell charge and Gauss-Bonnet terms, respectively. The background
is Reissner-Nordstr\"{o}m-AdS black brane solution in Gauss-Bonnet
gravity.  It is shown that these corrections affect drag force and
jet quenching parameter. We find an analytic solution of drag force
in this background which depends on Gauss-Bonnet coupling and
charge.  We set Gauss-Bonnet coupling to be zero and find drag force
in the case of Reissner-Nordstr\"{o}m-AdS background. Also we
discuss the relaxation time of a moving heavy quark in this gravity
background.
\end{abstract}

\vspace*{0.5cm}

 \maketitle
%\end{titlepage}
\section{Introduction}
The experiments of Relativistic Heavy Ion Collisions (RHIC) have
produced a strongly-coupled quark-gluon plasma
(QGP)\cite{Shuryak:2004cy}. The $AdS/CFT$ correspondence
\cite{Maldacena:1997re,Gubser:1998bc,Witten:1998qj, Witten:1998zw}
has yielded many important insights into the dynamics of
strongly-coupled gauge theories. It has been used to investigate
hydrodynamical transport quantities in various interesting
strongly-coupled gauge theories where perturbation theory is not
applicable. Methods based on $AdS/CFT$ relate gravity in $AdS_5$
space to the conformal field theory on the 4-dimensional boundary.
It was shown that an $AdS$ space with a black brane is dual to
conformal field theory at finite temperature. In the framework of
$AdS/CFT$, an external quark is represented as a string dangling
from the boundary of $AdS_5$-Schwarzschild and a dynamical quark is
represented as a string ending on flavor D7-brane and extending down
to some finite radius in $AdS$ black brane background. Then one can
calculate energy loss of quarks to the surrounding strongly-coupled
plasma.

 One of the interesting properties of the strongly-coupled
plasma at RHIC is jet quenching of partons produced with high
transverse momentum. The jet quenching parameter controls the
description of relativistic partons and it is possible to employ the
gauge/gravity duality and determine this quantity at the finite
temperature theories. There has been the $AdS/CFT$ calculation of
jet quenching parameter
\cite{Liu:2006ug,Buchel:2006bv,VazquezPoritz:2006ba,Caceres:2006as,
Lin:2006au,Avramis:2006ip,Armesto:2006zv,Argyres:2006vs,Argyres:2006yz,Argyres:2008eg}
and the drag coefficient which describes the energy loss for heavy
quarks in $\mathcal{N}=4$ supersymmetric Yang-Mills theory
\cite{Nakano:2006js, Herzog:2006gh, Herzog:2006se, Gubser:2006bz,
Gubser:2006qh, Caceres:2006dj, Matsuo:2006ws}.

The universality of the ratio of shear viscosity $\eta$ to entropy
density $s$
\cite{Policastro:2001yc,Kovtun:2003wp,Buchel:2003tz,Kovtun:2004de}
for all gauge theories with Einstein gravity dual raised the
tantalizing prospect of a connection between string theory and RHIC.
The results were obtained for a class of gauge theories whose
holographic duals are dictated by classical Einstein gravity. But
string theory contains higher derivative corrections from stringy or
quantum effects, such corrections correspond to $1/\lambda$ and
$1/N$ corrections. In the case of $\mathcal{N}=4$ super Yang-Mills
theory, the dual corresponds to type $\amalg B$ string theory on
$AdS_5\times S^5$ background. The leading order corrections in
$1/\lambda$ arises from stringy corrections to the low energy
effective action of type $\amalg B$ supergravity, $\alpha'^3 R^4$.

Recently, $\frac{\eta}{s}$ has been studied for a class of CFTs in
flat space with Gauss-Bonnet gravity
\cite{Brigante:2007nu,Brigante008gz,Kats:2007mq,Neupane:2008dc,Ge:2008ni,Buchel:2008wy}.
They compute the effect of  $R^2$ corrections to the gravitational
action in AdS space and show that the conjecture lower bound on the
$\frac{\eta}{s}$ can be violated. It was shown that in the
Reissner-Nordstr\"{o}m-AdS black brane solution in Gauss-Bonnet
gravity, $\frac{\eta}{s}$ bound is violated \cite{Ge:2008ni}. and
the Maxwell charge slightly reduces the deviation. Regarding this
study and motivated by the vastness of the string landscape
\cite{Douglas:2006es}, one can explore the modification of jet
quenching parameter and drag force on a moving heavy quark in the
strongly-coupled plasma. In general, we don't know about forms of
higher derivative corrections in string theory, but it is known that
due to string landscape one expects that generic corrections can
occur.

In this paper,  we investigate the corrections on jet quenching
parameter and drag force on a moving heavy quark in the Super
Yang-Mills plasma using the $AdS/CFT$ correspondence. We investigate
the effects of charge and finite 't Hooft coupling correction on
drag force and jet quenching parameter. The finite 't Hooft coupling
correction can be considered as stringy correction from $AdS/CFT$.
To study this correction, we consider Gauss-Bonnet terms. The effect
of the charge is considered by adding Maxwell charge. As it has been
proposed in \cite{Sin:2007ze}, one can consider charge either as the
R-charge or baryon charge. In the latter case, one can interpret
charge effect as the effect of finite baryon density. The background
is Reissner-Nordstr\"{o}m-AdS black brane solution in Gauss-Bonnet
gravity.  It is shown that these corrections affect drag force and
jet quenching parameter. We find an analytic solution of drag force
in this background which depends on Gauss-Bonnet coupling and charge.
We set Gauss-Bonnet coupling to be zero and find drag force in the case
of Reissner-Nordstr¨om-AdS background, too.%

In $\mathcal{N}=4$ SYM plasma, the gravity dual is type IIB
superstring theory and the inverse 't Hooft coupling correction to
the jet quenching parameter and drag force has been found in
\cite{Armesto:2006zv,Justin}. It is known that the curvature-squared
corrections are not the first higher derivative corrections in type
IIB superstring theory. These corrections on drag force have been
studied in \cite{Fadafan:2008gb}. The effects of finite-coupling
corrections and charge are calculated in (\ref{dragGBF}). We produce
the result of $\mathcal{N}=4$ super Yang-Mills plasma case in
(\ref{dragN4}). We also derive drag force in the case of
Reissner-Nordstr\"{o}m-AdS black brane solution. In this case, we
have considered only charge
effects to the drag force. \\%

The article is organized as follows. In section 2, we introduce the
action which describes the Gauss-Bonnet term and $U(1)$ gauge field.
Then we calculate drag force and study different limits of it. In
the next section we calculate jet quenching parameter and study the
effect of the corrections on this parameter. Finally, in the last
section we discuss our results together with possible extensions for
future work.

\section{Charge effect and finite coupling corrections on the drag force}

We study the gauge theory at finite temperature $T$ and assume the
geometry has a black hole.  In the gauge theory side, an external
quark can be introduced by a string that has a single end point at
the boundary and extends down to the horizon. We study the black
holes with higher derivative curvature in Anti-de Sitter space. In
five dimensions, the most general theory of gravity with quadratic
powers of curvature is Einstein-Gauss-Bonnet (EGB) theory. The exact
solutions and thermodynamic properties of the black branes in the
Gauss-Bonnet gravity have been discussed in
\cite{Cai:2001dz,Nojiri:2001aj,Nojiri:2002qn}. The authors in
\cite{Brigante:2007nu,Brigante008gz,Neupane:2008dc,Ge:2008ni} showed
that for a class of CFTs with Gauss-Bonnet gravity dual, the ratio
of shear viscosity to entropy density could violate the conjectured
viscosity bound. We try to understand more about the drag force on a
moving heavy quark in the boundary gauge theory by string trailing
in the Gauss-Bonnet gravity.

To study the effects of charge and finite 't Hooft coupling, we
consider the Reissner-Nordstr\"{o}m-AdS black brane solution in
Gauss-Bonnet gravity \cite{Cvetic:2001bk}. The following action in 5
dimensions describes the Gauss-Bonnet term and $U(1)$ gauge field
\begin{eqnarray}
S=\frac{1}{16\pi G_5}\int d^5 x \sqrt{-g}\,
[R-\Lambda+\lambda_{GB}\left( R^2-4 R_{\mu\nu}R^{\mu\nu}+
R_{\mu\nu\rho\sigma}R^{\mu\nu\rho\sigma}\right)-4\pi G_5
F_{\mu\nu}F^{\mu\nu}]\label{GBFaction},
\end{eqnarray}
where $\lambda_{GB}$ is Gauss-Bonnet coupling constant and the
negative cosmological constant is related to radius of AdS space by
$\Lambda=-\frac{12}{R^2}$. We also know from the standard Maxwell
action that the field strength is given by
$F_{\mu\nu}=\partial_{\mu}A_{\nu}-\partial_{\nu}A_{\mu}$. The
charged black brane solution in 5-dim is
\begin{equation}
ds^2=-N^2 \,\frac{r^2}{R^2}\, f(r)\,
dt^2+\frac{dr^2}{\frac{r^2}{R^2} f(r)}+\frac{r^2}{R^2}
\,d\vec{x}^2\label{GBFmetric},
\end{equation}
where
\begin{equation}
f(r)= \frac{1}{2\lambda_{GB}}\left( 1-\sqrt{1-4 \lambda_{GB}\left(
1-\frac{m R^2}{r^4}+\frac{q^2\,R^2}{r^6} \right)}\right).
\end{equation}
In our coordinates, $r$ denotes the radial coordinate of the black
hole geometry and $t, \vec{x}$ label the directions along the
boundary at the spatial infinity. In these coordinates the event
horizon is located at $f(r_h)=0$ where $r_h=r_+$ is the largest root
and it can be found by solving this equation. The boundary is
located at infinity and the geometry is asymptotically AdS with
radius R. The electric field strength for the Maxwell charge $q$ is
$ F\equiv\frac{q^2}{4\pi\,r^6}dt\wedge\,dr$. The gravitational mass
of $M$ and the charge $Q$ are expressed as $
M=\frac{3\,V_{3}R^2\,m}{16\pi G_5},\,\,\,\,\,\, Q=\frac{q^2}{12\pi }
$. The constant $N^2$ is arbitrary which specifies the speed of
light of the boundary gauge theory and we choose it to be unity. As
a result at the boundary, where $r\rightarrow\infty$,
\begin{equation}
f(r)\rightarrow \frac{1}{N^2 }, \,\,\,\,\, N^2= \frac{1}{2}\left(
1+\sqrt{1-4 \lambda_{GB}} \right)\label{a}.
\end{equation}
Beyond  $\lambda_{GB}\leq\frac{1}{4}$ there is no vacuum AdS
solution and one cannot have a conformal field theory at the
boundary. Casuality leads to new bounds on $\lambda_{GB}$. The
authors in \cite{Brigante008gz,Ge:2009eh} found that in five
dimensions $\lambda_{GB}$ is less than $.09$ and when dimensions of
space-time go up, casuality restricts the value of $\lambda_{GB}$ in
the region $\lambda\leq 0.25$.%

The temperature of the hot plasma is given by the Hawking
temperature of the black hole
\begin{equation}
T=\frac{N r_h}{2 \pi R^2} \left( 2-\frac{q^2
R^2}{r_h^2}\right)=\frac{N r_h}{2 \pi R^2} \left( 2-\tilde{q}
\right)=\frac{1}{2}\,T_0\,N\,\left( 2-\tilde{q} \right)
\label{TGBF}.
\end{equation}
where $\tilde{q}=\frac{q^2 R^2}{r_h^2}$ and $T_0$ is the temperature
of AdS black hole solution without any corrections. In this limit
$\tilde{q}\rightarrow 2$, the black brane approaches to the extremal
case. We introduce a new dimensionless coordinate
$u=\frac{r_h^2}{r^2}$, the five-dimensional metric (\ref{GBFmetric})
is  deformed as
\begin{equation}
ds^2=-\frac{N^2\,f(u)\,r_h^2}{R^2\,u}\, dt^2+\frac{R^2\,du^2}{4 u^2
f(u)}+\frac{r_h^2\,}{R^2\,u} \,d\vec{x}^2\label{GBFnewmetric},
\end{equation}
where
\begin{equation}
f(u)= \frac{1}{2\lambda_{GB}}\left( 1-\sqrt{1-4 \lambda_{GB}\left(
1-u\right)\left(1+u-\tilde{q}\,u^2\right)}\right).\label{fu}
\end{equation}
Now in this coordinate, the event horizon is located at $u=1$, while
$u=0$ is located at the boundary of the AdS space.

The relevant string dynamics is captured by the Nambu-Goto action
\begin{eqnarray}
S=-\frac{1}{2 \pi \alpha'}\int d\tau d\sigma\sqrt{-det\,g_{ab }
},\label{Numbo-Goto}
\end{eqnarray}
where the coordinates $(\sigma, \tau)$ parameterize the induced
metric $g_{ab}$ on the string world-sheet and $X^\mu(\sigma, \tau)$
is a map from the string world-sheet into the space-time. Defining
$\dot X = \partial_\tau X$, $X' =
\partial_\sigma X$, and $V \cdot W = V^\mu W^\nu G_{\mu\nu}$ where
$G_{\mu\nu}$ is the AdS black brane metric (\ref{GBFmetric}), we
have
\begin{equation}
-det\,g_{ab }=(\dot X \cdot X')^2 - (X')^2(\dot X)^2.
\end{equation}
One can make the static choice  $\sigma=u, \tau=t$ and following
\cite{Herzog:2006gh,Gubser:2006bz} focus on the dual configuration
of the external quark moving in the $x^1$ direction on the plasma.
The string in this case, trails behind its boundary endpoint as it
moves at constant speed $v$ in the $x^1$ direction
\begin{equation}
x^1(u,t)=v t+\xi(u),\,\,\,\,\,  x^2=0,\,\,\, x^3=0.
\end{equation}
given this, one can find the lagrangian as follows
\begin{equation}
\mathcal{L}=\sqrt{-det\,g_{ab}}=r_h\,\sqrt{\frac{N^2}{4
u^3}-\frac{v^2}{4
 u^3
f(u)}+\frac{f(u)\,N^2\,r_h^2}{R^4\,u^2}\,\xi'^2},\label{lagGBF}
\end{equation}
The equation of motion for $\xi$  implies that $\frac{\partial
L}{\partial \xi'}$ is a constant. One can name this constant as
$\Pi_{\xi}$ and it can be found from (\ref{lagGBF}). We solve the
relation for $\xi'$, the result is
\begin{equation}
\xi'^2=\frac{R^4\,\Pi_\xi^2}{4\,u\,f(u)\,N^2 r_h^2}\left(\frac{
N^2-\frac{v^2}{f(u)}}{ \frac{f(u)\,N^2\,r_h^4}{R^4\,u^2}- \Pi_\xi^2
}\right).\label{XiGBF}
\end{equation}
We are interested in a string that stretches from the boundary to
the horizon. In such a string, $\xi'^2$ remains positive everywhere
on the string. Hence both numerator and denominator change sign at
the same point and with this condition, one can find the constant of
motion $\Pi_{\xi}$ in terms of the critical value of $u_c$ as follows%
\beq \Pi_{\xi}=\frac{v\,r_h^2}{R^2\,u_c}, \eeq%

Numerator and denominator in (\ref{XiGBF}) change sign at $u_c$ and it can be found by solving this equation%
\beq \frac{1}{2\lambda_{GB}}\left( 1-\sqrt{1-4 \lambda_{GB}\left(
1-u_c\right)\left(1+u_c-\tilde{q}\,u_c^2\right)}\right)-\frac{v^2}{N^2}=0.
\eeq

The drag force that is experienced by the heavy quark is calculated
by the current density for momentum along $x^1$ direction. After
straightforward calculations, the drag force is easily simplified in
terms of $\Pi_{\xi}$
\begin{equation}
F=-\frac{1}{2 \pi \alpha'} \Pi_\xi\label{generaldrag}.
\end{equation}

Plugging these relations into the drag force leads to the following
result
\begin{equation}
F\left(\lambda_{GB},q\right)=-\frac{r_h^2}{2 \pi \alpha'\,R^2}\left(
\frac{v\,(24\,N^4)^{\frac{1}{3}}\,
\left(N^4-N^2\,v^2+\lambda_{GB}\,v^4 \right) \,(1+\tilde{q})
+2^{\frac{1}{3}}\,v\,\mathbf{y}^{\frac{2}{3}}}{6^{\frac{2}{3}\,}N^{\frac{4}{3}}
\,\left(N^4-N^2\,v^2+\lambda_{GB}\,v^4
\right)\,\mathbf{y}^{\frac{1}{3}}}\right)\,,\label{dragGBF}
\end{equation}
where $\mathbf{y}$ is a function of coupling constant of
Gauss-Bonnet, charge and velocity of heavy quark as
\begin{eqnarray}
\mathbf{y}&=&-9\,\tilde{q}\,\left(N^4-N^2\,v^2+\lambda_{GB}\,v^4
\right)^2\,\nonumber\\
&&+\sqrt{3}\,\sqrt{\left(N^4-N^2\,v^2+\lambda_{GB}\,v^4
\right)^3\,\left(27\,\tilde{q}^2\,\lambda_{GB}\,v^4-27\,
\tilde{q}^2\,N^2\,v^2-\left(\tilde{q}-2\right)^2(1+4\,\tilde{q})N^4\right)}\,.\nonumber
\\
\end{eqnarray}
As sequence, drag force depends on the coupling constant of
Gauss-Bonnet $\lambda_{GB}$ and Maxwell charge $q$. One can express
the drag force in (\ref{dragGBF}) in terms of gauge theory
parameters. Notice that $\tilde{q} = \frac{R^2}{r^2_h}\, q^2$ then
one should use (\ref{TGBF}) to write $r_h$ in terms of the
temperature of the hot plasma.

It would be interesting to discuss the extremal case, too. For
$\tilde{q} = 2$, the temperature of the plasma is zero and the black
hole becomes an extremal one with a double zero at the horizon. We
note that in the extremal case, $\tilde{q}=2$ the drag force is
regular \cite{progress}.

\subsection{special limits of drag force}
In this section, we will discuss different limits of drag force.
Drag force in these limits can be calculated directly from the
related action. Here, we derive results from the drag force in
(\ref{dragGBF}).

First of all, let us consider the case of $\lambda_{GB}\rightarrow0$
and $q\rightarrow0$ in (\ref{dragGBF}). In this limit, one does not
consider any correction in the action (\ref{GBFaction}). The drag
force $F\left(\lambda_{GB}\rightarrow0,q\rightarrow0\right)$ is
nothing but the drag force in the case of $\mathcal{N}=4$
strongly-coupled SYM plasma $F_{\mathcal{N}=4}$. The authors of
\cite{Gubser:2006bz,Herzog:2006gh} have obtained
\begin{equation}
F_{\mathcal{N}=4}=-\left(\frac{\pi\,\sqrt{\lambda}\,T_0^2}{2}\right)\,\frac{v}{\sqrt{1-v^2}}.\label{dragN4}
\end{equation}
where $\lambda$ is 't Hooft coupling.

Now, we will investigate the drag force in two following special
limits.\\

\begin{itemize}
\item{
In the case of $\,\,\lambda_{GB}\neq0$, $q\rightarrow0\,.$ \\\\
In this limit, the action (\ref{GBFaction}) is reduced to the
Gauss-Bonnet action without Maxwell action. It corresponds to the
effect of finite-coupling corrections to the drag force on a moving
heavy quark in the Super Yang-Mills plasma. These corrections are
related to curvature-squared corrections in the corresponding
gravity dual \cite{Fadafan:2008gb,Justin}. We name this case as
$F\left(\lambda_{GB}\right)=F\left(\lambda_{GB},q\rightarrow0\right)$.
One can find drag force in this limit from (\ref{dragGBF}) and
derive this result
\begin{equation}
F\left(\lambda_{GB}\right)=-\left(\frac{\pi\,\sqrt{\lambda}}{2}\,T_{GB}^2\right)
\frac{v}{\sqrt{N^4-N^2\,v^2+\lambda_{GB}\,v^4}}.\label{FGB}
 \end{equation}

where $T_{GB}$ is the Hawking temperature of Gauss-Bonnet black hole
and is defined as%
\beq T_{GB}=\frac{N\,r_h}{\pi\,R^2}. \eeq

As it is seen from Eq. (\ref{FGB}), the drag force depends on the
Gauss-Bonnet coupling constant. It is worse to mention that our
result is exactly the same as ones which are obtained in
\cite{Fadafan:2008gb,Justin}.}
\item{
In the case of $\,\,q \neq 0$ and $ \lambda_{GB} \rightarrow0\,.$\\\\
In this case, we consider only the Maxwell action in
(\ref{GBFaction}) where it has the Reissner-Nordstr\"{o}m-AdS black
brane solution.  This is the first calculation of drag force in this
background. One can follow the definition of drag force in
(\ref{generaldrag}) and derive the following result from the
Reissner-Nordstr\"{o}m-AdS black brane solution.

Applying this limit, drag force is found as
\begin{eqnarray}
&&F(\tilde{q})=-\frac{r_h^2}{2 \pi \alpha'\,R^2}\,v \nonumber\\
&&\times\frac{24^{\frac{1}{3}}\left(1-v^2\right)\left(1+\tilde{q}\right)
+ 2^{\frac{1}{3}}\left( \sqrt{3\,\left(v^2-1\right)^3
\left(\left(1+4 \tilde{q}\right)\left(\tilde{q}-2\right)^2+27
\tilde{q}^2 v^2\right)}-9 \tilde{q} (1-v^2)^2\right)^{\frac{2}{3}}}
{6^{\frac{2}{3}}\left(1-v^2\right)\left(
\sqrt{3\,\left(v^2-1\right)^3 \left(\left(1+4
\tilde{q}\right)\left(\tilde{q}-2\right)^2+27 \tilde{q}^2
v^2\right)}-9
\tilde{q}(1-v^2)^2\right)^{\frac{1}{3}}}.\label{dragmedium}\nonumber\\
\end{eqnarray}}
The Hawking temperature of Reissner-Nordstr\"{o}m-AdS black brane is
given by%
\beq T_q=\frac{r_h}{2\,\pi R^2}\left( 2-\tilde{q} \right).
\label{Tq}\eeq
\end{itemize}
One can write the drag force in (\ref{dragmedium}) in terms of gauge
theory parameters. Notice that $\tilde{q} = \frac{R^2}{r^2_h}\, q^2$
and one should use (\ref{Tq}) to write $r_h$ in terms of the
temperature of the plasma.

\section{Jet Quenching parameter}
In this section, we calculate the "jet quenching parameter"
$\hat{q}$. This quantity provides a measure of the dissipation of
the plasma. Because the quark gluon plasma at RHIC is a
strongly-coupled plasma, one expects that an energetic quark looses
its energy very fast. This is related to the main point that back to
back jets are suppressed.

In \cite{Liu:2006ug}, Liu, Rajagopal and Wiedemann proposed a first
principle, nonperturbative quantum field theoretic definition of
$\hat{q}$. Following their proposal, we consider a string whose end
points lie at $r\rightarrow\infty$ in the space time
(\ref{GBFmetric}) and trace a rectangular light-like Wilson loop.
One should consider metric (\ref{GBFnewmetric}) in the light-cone
coordinates
\begin{equation}
x^{\pm}=\frac{1}{\sqrt{2}}(x^0\pm x^1).
\end{equation}
The relevant Wilson loop is a rectangle with large extension $L^-$
in the $x^-$ direction and small extension $L$ along the $x^2$
direction. Then, $\hat{q}$ is given by the large $L^+$ behavior of
the Wilson loop
\begin{equation}
W = e^{-\frac{1}{4\sqrt{2}}\hat{q}\,L^-L^2}.
\end{equation}

We choose the static gauge in which
\begin{equation}
\tau=x^-\,,\,\sigma=x^2.
\end{equation}
By assuming a profile of $u=u(\sigma)$ and substituting induced
metric of fundamental string into the Nambu-Goto action
(\ref{Numbo-Goto}), one obtains
\begin{equation}
S=\frac{L^-}{\sqrt{2}\pi\alpha'}\int_0^{\frac{L}{2}}d\sigma \,
\frac{r_h}{R\,u}\,\sqrt{\left(1-f(u)N^2\right)\left(\frac{r_h^2}{R^2
}-\frac{u'^2R^2}{4 u f(u)}\right)}\,.\label{S}
\end{equation}
where $u'=\partial_{\sigma}\,u$. Regarding $\sigma$ as time, this
point that the lagrangian is independent of time implies that the
energy is conserved. The boundary condition for bulk coordinate $r$
is
\begin{equation}
r\left(\pm \frac{L}{2}\right)=\infty\,.
\end{equation}
and such an embedding preserves  a symmetry $r(\sigma)=r(-\sigma)$.
This symmetry implies that $r'(\sigma=0)=0$ and one concludes that
at $\sigma=0$ the string world sheet must reach the horizon,
$r(\sigma=0)=r_0$. Now from (\ref{S}), one can find the transverse
width of Wilson loop, $L$.\\

We must subtract from action $S$, the self-interaction of isolated
quark and antiquark, which corresponds to the Nambu-Goto action of
two strings that extend from boundary $(u\rightarrow 0)$ to the
horizon $(u=1)$. This action is $S_0$ and one can calculate it. The
resulting $S_I=S-S_0$ is the extremal action in the definition of
jet quenching parameter in \cite{Liu:2006ug}. There are some points
about the extremal action in \cite{Argyres:2006yz,Argyres:2008eg}.

We follow the approach \cite{Liu:2006ug} and calculate jet quenching
parameter. In this way, we are interested in the expectation value
of thermal Wilson loop as $L\rightarrow 0$.  After applying this
limit on $S$, the leading order term is canceled by subtracting
$S_0$ in $S_I$. Considering metric in (\ref{GBFnewmetric}), $S_I$
can be found as the following
\begin{equation}
S_I=\frac{r_h^3 \,L^- L^2}{4\,\sqrt{2} \,R^4\, \pi
\alpha'}\,I^{-1}\,,\label{SI}
\end{equation}
where $I$ is
\begin{equation}
I=\int_0^1 \,du\,\sqrt{\frac{u}{f(u)\left(1-f(u) N^2\right)}}\,.
\end{equation}
and $f(u)$ was defined in (\ref{fu}).

Using Eq. (\ref{SI}) and the relation
$\alpha'=\frac{R^2}{\sqrt{\lambda}}$, the jet quenching parameter is
\begin{equation}
\hat{q} =\frac{2\,r_h^3\,\sqrt{\lambda}}{R^6\,\pi}\,I^{-1}\,.
\label{Qh}\end{equation} Let us study the behavior of jet quenching
parameter in terms of Gauss-Bonnet coupling constant and charge
corrections.

First, we consider there are no charges in (\ref{Qh}). In this case, $I_{GB}$ is defined as%
\beq I_{GB}=\int_0^1 \,du\,\sqrt{\frac{u}{h(u)\left(1-h(u)
N^2\right)}},\eeq
where
\beq h(u)=\frac{1}{2\lambda_{GB}}\left(1-\sqrt{1-4
\lambda_{GB}(1-u^2)}\right). \eeq
Using the above equation, we have plotted $I^{-1}_{GB}$ versus
$\lambda_{GB}$ in the left plot of figure 1. Based on this result,
one finds that the jet quenching parameter is enhanced due to the
Gauss-Bonnet corrections with positive $\lambda_{GB}$, while
$\hat{q}$ decreases with negative $\lambda_{GB}$. It is worse to
mention that our results are in good agreement with
\cite{Argyres:2008eg}.

Now, we discuss the jet quenching parameter at the presence of
finite charge and no Gauss-Bonnet corrections. In this case, one finds
\beq I_{charge}=\int_0^1 \,du\,\sqrt{\frac{u}{g(u)\left(1-g(u)
\right)}},\label{Icharge}\eeq
where
\beq g(u)=(1-u)\left(1+u-\tilde{q}\, u^2\right). \eeq

We have also plotted $I^{-1}_{charge}$ versus $\tilde{q}$ in the
right plot of  figure 1. Based on this result, one can find the jet
quenching parameter. As it is obviously seen from this plot, there
is a maximum value for $I^{-1}_{charge}$ at $\tilde{q}_c=0.750$.
This critical value of charge can be found by studying the
derivative of (\ref{Icharge}). In other words, $I^{-1}_{charge}$ is
increased with $\tilde{q}$ till $\tilde{q}_c$ and then it is
decreased and goes to zero at maximum value of charge $\tilde{q}=2.$

If we set charge and Gauss-Bonnet coupling corrections to be zero,
one finds that %
\beq I_{0}=\int_0^1 \,du\,\sqrt{\frac{1}{u\left(1-u^2
\right)}}=\frac{2\,\pi^{\frac{1}{2}}\,\Gamma[\frac{5}{4}]}{\Gamma[\,\frac{3}{4}]}\,. \eeq%
This case is related to $\mathcal{N}=4$ SYM theory. It is
straightforward to derive Jet quenching parameter as
\begin{equation}
\hat{q}_{\mathcal{N}=4}=\frac{\pi^{\frac{3}{2}}\,\Gamma[\frac{3}{4}]}{\Gamma[\,\frac{5}{4}]}\sqrt{\lambda}\,T_0^3\,.\label{qhat}
\end{equation}
where $T_0$ is the temperature of the SYM $\mathcal{N}=4$ plasma.
This result is exactly the same as the result in \cite{Liu:2006ug}.
%%%%%%%%%%%
\begin{figure}
  \centerline{\includegraphics[width=3in]{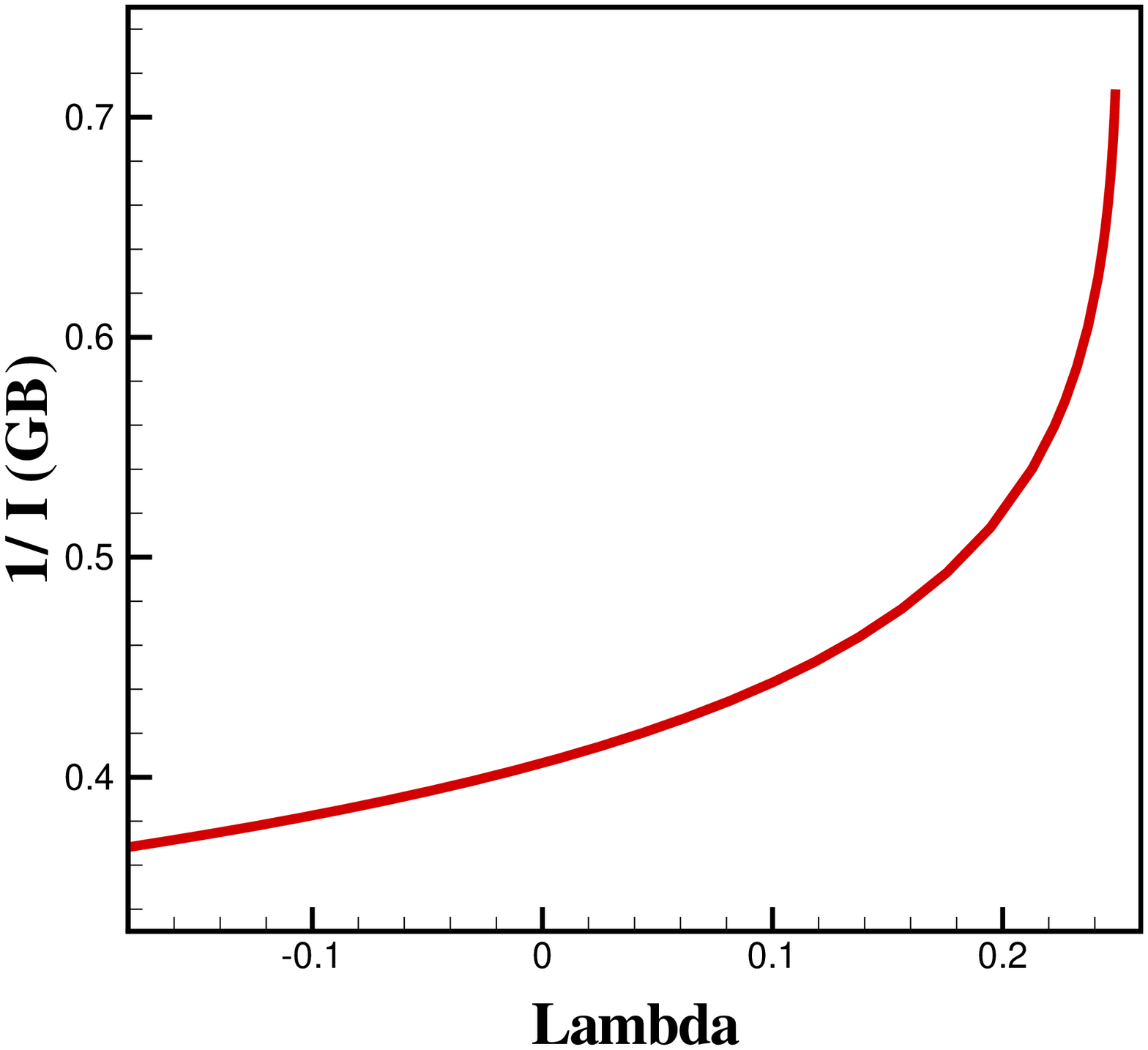} \includegraphics[width=3in]{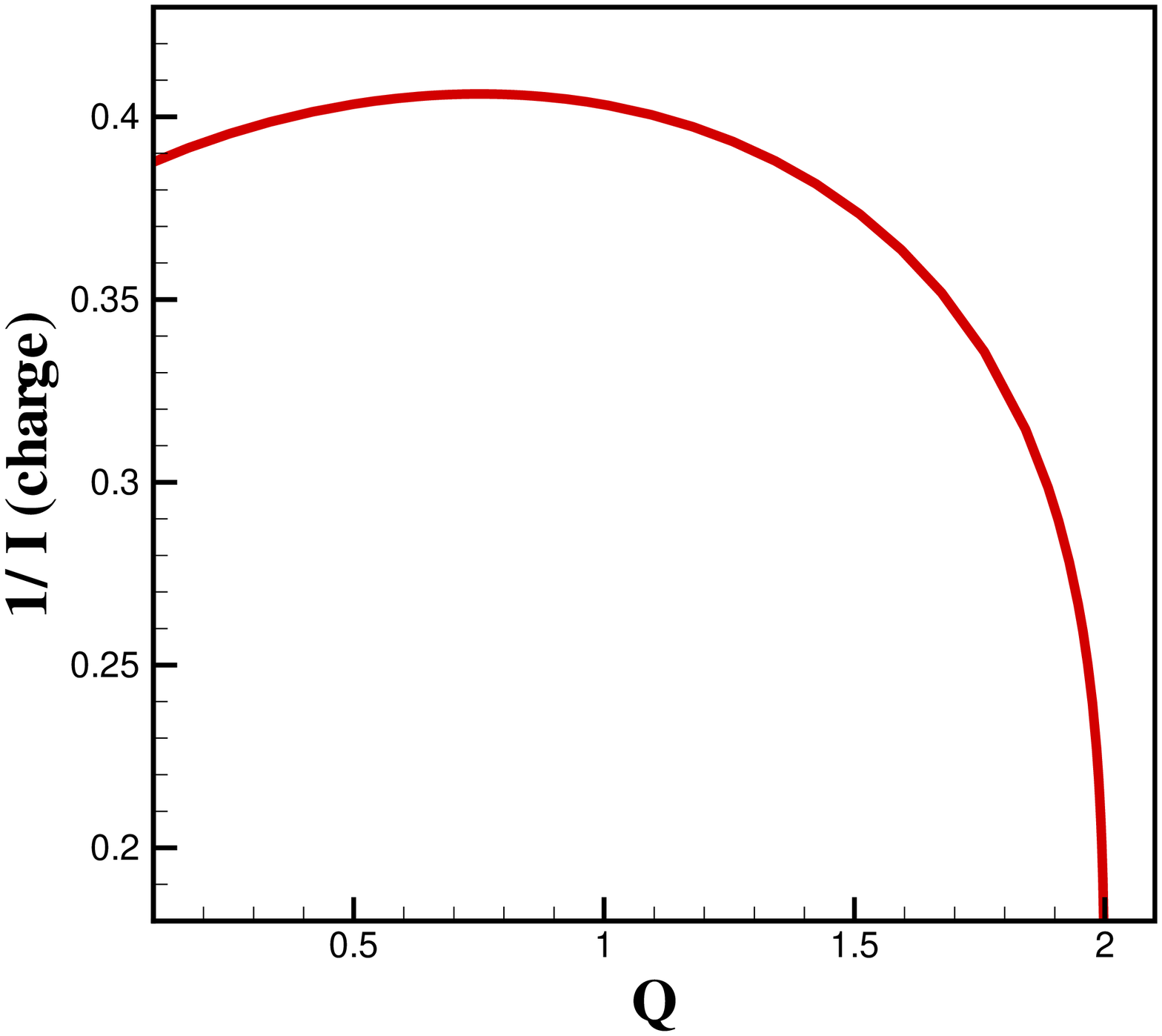}}
  \caption{The left figure shows $I^{-1}_{GB}$ versus
  $\lambda_{GB}$. In this figure, we consider only Gauss-Bonnet
  corrections. The figure on the right shows
  $I^{-1}_{charge}$ versus $\tilde{q}$. In this case we consider only the effect of the charge.}\label{fig}
 \end{figure}
%%%%%%%%

One can discuss the jet quenching parameter at different
temperatures in the experimentally relevant range. In this case,
there are different temperature matching schemes in
\cite{Gubser:2006qh}.

\section{Discussion}

In this paper, we have computed the drag force on a moving heavy
quark and the jet quenching parameter in a field theory dual to
gravity with Gauss-Bonnet terms. In addition to previous
publications here a Maxwell charge has been added. Also the jet
quenching parameter in this background is new.

Recently, $\frac{\eta}{s}$ has been studied for a class of CFTs in
flat space with Gauss-Bonnet gravity
\cite{Brigante:2007nu,Brigante008gz,Kats:2007mq,Neupane:2008dc,Ge:2008ni,Buchel:2008wy}.
They compute the effect of  $R^2$ corrections to the gravitational
action in AdS space and show that the conjecture lower bound on the
$\frac{\eta}{s}$ can be violated. Also it was shown that in the
Reissner-Nordstr\"{o}m-AdS black brane solution in Gauss-Bonnet
gravity, $\frac{\eta}{s}$ bound is violated \cite{Ge:2008ni}.
Actually the Maxwell charge slightly reduces the deviation. As long
as the universality of $\frac{\eta}{s}$ bound is not settled, a
detailed phenomenology of theories with gravity dual is interesting
and might reveal the common conceptual reasons for violating or
obeying the viscosity bound.

In this study, we have considered two corrections on the drag force
and jet quenching parameter; finite coupling correction and charge
effect. To study finite coupling correction, we considered
Gauss-Bonnet terms. The effect of the charge is considered by adding
Maxwell charge where one can interpret charge effect as the effect
of finite baryon density \cite{Sin:2007ze}. The background is
Reissner-Nordstr\"{o}m-AdS black brane solution in Gauss-Bonnet
gravity.

It is shown that these corrections affect drag force and jet
quenching parameter. We found an analytic solution of drag force in
this background which depends on Gauss-Bonnet coupling and charge.
We also derived drag force in the case of Reissner-Nordstr\"{o}m-AdS
black brane solution. In this case, we have considered only charge
effects to the drag force.

From phenomenology point of view, we discuss relaxation time and
diffusion coefficient of a non-relativistic heavy quark. Relaxation
time and diffusion coefficient of a non-relativistic heavy quark
have been studied in
\cite{Herzog:2006gh,Gubser:2006qh,CasalderreySolana:2006rq}. The
effect of curvature-squared corrections on these quantities has been
studied in \cite{Fadafan:2008gb}. We discuss the finite coupling
corrections on thermal transport properties of a moving heavy quark
in gauss Bonnet gravity. The drag force has been calculated in
(\ref{GBdrag}), we use this expression and find relaxation time and
diffusion coefficient of a non-relativistic heavy quark.

The drag force in the Gauss-Bonnet background is given by
\begin{equation}
F\left(\lambda_{GB}\right)=-\frac{\pi \sqrt{\lambda }}{2}T_{GB}^2
\frac{v}{\sqrt{N^2(N^2-v^2)+\lambda_{GB}v^4}}\,.\label{GBdrag}
\end{equation}
we follow the approach in \cite{Fadafan:2008gb}. Because of the
small velocity of non-relativistic heavy quark, one can neglect the
squared-velocity term in (\ref{GBdrag}) and momentum of heavy quark
is given by $p=m\,v$. It is clear the momentum of heavy quark fall
off exponentially
\begin{equation}
p(t)=p(0)\,e^{-\frac{t}{\,t_{\lambda_{GB}}}},\,\,\,\,\,\,\,\,\,\,\,\,\,t_{\lambda_{GB}}=\frac{
 m\,\left(1+\sqrt{1-4\,\lambda_{GB}}\right)}{\pi \sqrt{\lambda }\,T_{GB}^2}
\end{equation}
where $t_{\lambda_{GB}}$ is relaxation time of heavy quark in
Gauss-Bonnet gravity.

One can derive the result of $\mathcal{N}=4$ case by considering
$\lambda_{GB}=0$
\begin{equation}
t_{\mathcal{N}=4}=\frac{2\,m}{\pi\,T_0^2\,
\sqrt{\lambda}}\label{DN4}\,.
 \end{equation}
This result has been discussed in
\cite{Herzog:2006gh,Gubser:2006qh}.

The diffusion coefficient is related to the temperature of the
plasma $T$, the heavy quark mass $m$ and the relaxation time $t_D$
as $D=\frac{T}{m}\,t_D$.  It is straightforward to obtain the
diffusion coefficient in the case of $\mathcal{N}=4$ SYM plasma
\cite{Herzog:2006gh,Gubser:2006qh}
\begin{equation}
D_{\mathcal{N}=4}=\frac{2}{\pi\,T\, \sqrt{g_{YM}^2 N}}\label{DN4}\,.
\end{equation}
This result has been achieved with a different approach in
\cite{CasalderreySolana:2006rq}, for non-relativistic heavy quarks.
Using the above approach, one can obtain the diffusion coefficient
of a moving heavy quark in Gauss-Bonnet gravity, too.
Also the extremal case can be considered to calculate drag force and
jet quenching parameter. To clarify this case, one finds from Eq.
(\ref{TGBF}) that $\tilde{q} = 2$ corresponds to $T = 0$. As a
result, one should consider a moving heavy quark in finite charge
plasma at zero temperature. In this case drag force is finite
\cite{progress}.

\acknowledgments{I would like to thank J. Abouie for useful
discussions and specially thank the referee for his/her comments.}\\
%%%%%%%%%%%%%%%%%%%%%%%%%%%%%%%%%%%%%%%%%%%%%%%%%%%% refrences %%%%%%%%%%%%%%%%%%%%%%%%%
%%%%%%%%%%%%%%%%%%%%%%%%%%%%%%%%%%%%%%%%%%%%%%%%%%%%%%%%%%%%%%%%%%%%%%%%%%%%%%%%%%%%%%
%%%%%%%%%%%%%%%%%%%%%%%%%%%%%%%%%%%%%%%%%%%%%%%%%%%%%%%%%%%%%%%%%%%%%%%%%%%%%%%%%

\end{document}